\documentclass[%
 preprint,
 showkeys,
 amsmath,amssymb,
 aps,
]{revtex4-1}

\usepackage[ansinew]{inputenc}% Include German Umlaute
\usepackage{graphicx}% Include figure files
\usepackage{dcolumn}% Align table columns on decimal point
\usepackage{bm}% bold math
\usepackage{hyperref}% add hypertext capabilities
\usepackage[mathlines]{lineno}% Enable numbering of text and display math
\begin{document}

\preprint{APS/123-QED}

\title{Spectroscopy of the $D_1$-transition of cesium by dressed-state resonance fluorescence from a single (In,Ga)As/GaAs quantum dot}

\author{S. M. Ulrich$^{1,\dag}$\email{Corr. author: s.ulrich@ihfg.uni-stuttgart.de}, S. Weiler$^{1}$, M. Oster$^{1}$, M. Jetter$^{1}$, A. Urvoy$^{2}$, R. Löw$^{2}$, and P. Michler$^{1}$}

\affiliation{$^1$~Institut für Halbleiteroptik und Funktionelle Grenzflächen, Universität Stuttgart, Allmandring 3, 70569 Stuttgart, Germany}
\affiliation{$^2$~5. Physikalisches Institut, Universität Stuttgart, Pfaffenwaldring 57, 70569 Stuttgart, Germany.}

%\collaboration{MUSO Collaboration}
%\noaffiliation

\date{\today}% Default is 'today', but any date may be explicitly specified.

\begin{abstract}
We use a laser-driven single (In,Ga)As quantum dot (QD) in the dressed state
regime of resonance fluorescence ($T = 4$~K) to observe the four $D_1$-transition
lines of alkali atomic cesium ($Cs$) vapor at room temperature. We tune the frequency of the dressing continuous-wave
laser in the vicinity of the bare QD resonance $\sim 335.116$~THz ($\sim 894.592$~nm)
at constant excitation power and thereby controllably tune the center and side channel frequencies of
the probe light, i.e. the Mollow triplet. Resonances between individual QD Mollow triplet lines
and the atomic hyperfine-split transitions are clearly identified in the $Cs$ absorption spectrum.
Our results show that narrow-band (In,Ga)As QD resonance fluorescence (RF) is suitable to optically
address individual transitions of the $D_1$ quadruplet
without applying magnetic field or electric field tuning.

\begin{description}

\item[PACS numbers]
78.55.Cr, 78.67.Hc, 78.90.+t, 32.10.Fn, 32.80.Xx, 32.90.+a
% III-V semiconductors, Quantum dots, Other topics in optical properties / condensed matter spectroscopy and other interactions of particles and radiation with condensed matter,
% Fine and hyperfine structure, Level crossing and optical pumping, Other topics in atomic properties and interactions of atoms with photons

\end{description}
\end{abstract}

\pacs{Valid PACS appear here}% PACS, the Physics and Astronomy Classification Scheme.

\keywords{Semiconductor, quantum dot, resonance fluorescence, alkali atom, cesium, hyperfine structure} % Use showkeys class option if keyword display desired

\maketitle

Hybrid quantum systems, particularly combining the characteristic strengths of semiconductor quantum dots (QDs) and alkali atomic vapors, have recently attracted
tremendous technological interest in view of sophisticated applications and future quantum logic devices in the field of quantum information
processing. For example, very recent studies have demonstrated a first hybrid interface between a semiconductor ''artificial atom'' (GaAs QD) and
alkali atomic \textit{rubidium} ($^{87}Rb$) vapor, where the 6.8~GHz ($28\,\mu$eV) split $D_2$ double resonance of $Rb$ served as a strongly
dispersive medium for QD luminescence photons. In these studies, spectral QD-$D_2$ resonance between both systems was achieved by either
electrical~\cite{Akopian.ea:2010} or magnetic field tuning~\cite{Akopian.ea:2011}, enabling slow light
propagation. In the latter case, photon storage by up to 15~times the temporal width of the wave packets could be achieved.
Furthermore, an energy stabilization scheme based on the absorption of QD photons in the $Rb$ vapor in combination with an active feed-back loop
for compensation of local carrier-induced spectral QD emission fluctuations was implemented by the same group~\cite{Akopian.ea:2010,Akopian.ea.arXiv:2013}.
Though, all of these previous studies were based on conventional, non-resonant schemes of optical excitation into the barrier matrix,
which results in inhomogeneously broadened QD emission lines (FWHM~$\approx\,80\,\mu\text{eV}\,\approx 19\,\text{GHz}$),
i.e. much larger than the corresponding atomic transition band widths and/or hyperfine structure splitting. The latter
aspect represents a strongly limiting factor of the principally achievable spectral resolution on one hand, but especially
will also strongly restrict or even hinder the desired level of high efficiency in hybrid functionality between semiconductor
QDs and atoms in future applications on the other hand.

Here we demonstrate an alternative hybrid approach to achieve controlled interaction between a
single semiconductor (In,Ga)As/GaAs QD and the $D_1$ transitions of alkali atomic \textit{cesium} vapor ($^{133}Cs$)
with high spectral resolution. In our scheme, we use ''dressed state'' resonance fluorescence (RF) from a single
(In,Ga)As QD under continuous-wave s-shell excitation~\cite{Muller.ea:2007,Flagg.ea:2009,Ates.ea:2009} to prepare a
quantum light source of narrow-band photon emission~\cite{Ates.ea:2009,Ulrich.ea:2011} approaching the Fourier transform limit,
provided by the individual components of the characteristic ''Mollow triplet''~\cite{Mollow:1969}.
The use of such a ''dressed'' quantum light source is of particular interest, as it allows to spectrally tune, i.e. control
the photon emission frequencies over a wide range of typically a few GHz~\cite{Vamivakas.ea:2009,Ulhaq.ea:2012,Ulhaq.ea:2013}.
As will be shown in the following, the use of well-separated individual components of the Mollow triplet
provides high spectral resolution of $\Delta \nu \sim 1$~GHz. This is comparable to the Doppler width ($\sim 0.5$~GHz)
of the atomic $Cs$ transitions \cite{Siddons.ea:2008} and allows to resolve the hyperfine-split $D_1$
quadruplet of the $6^2P_{1/2} \leftrightarrow 6^2S_{1/2}$ transitions in absorption.

The investigated sample structure was grown by metal-organic vapor-phase epitaxy. It is based on a single
layer of self-assembled (In,Ga)As/GaAs QDs emitting at 885-900~nm which are embedded in a planar wave\-guide structure, consisting
of a GaAs $\lambda$-cavity between 29 (4) periods of $\lambda/4$-thick AlAs/GaAs layers as bottom
(top) distributed Bragg reflectors (DBRs). The sample was kept at a temperature of $T = 4.0\,\pm\,0.5$~K
and excited by a narrow-band (FWHM~$\sim\,500$~kHz) tunable continuous-wave (cw) Ti:Sapphire laser.
Filtering and detection of the QD emission was performed by a grating monochromator (1200~l/mm) equipped with a CCD camera,
or alternatively, by a sensitive avalanche photo diode (APD). Combined with a 50:50 beam splitter in the detection path, two of such APDs served also as the
''start/stop'' triggers to determine the emission statistics by $g^{(2)}(\tau)$-type photon correlation measurements
(temporal resolution: $\Delta \tau_{det} \approx 400$~ps). Efficient suppression of parasitic laser stray light was achieved
by using an orthogonal geometry between lateral waveguide excitation into the cleaved sample side facet
and detection perpendicular to the DBR surface, combined with spatial filtering (pin-hole) and polarization suppression within the detection path.
To examine the dressed emission spectra of individual, resonantly excited QDs, high-resolution photoluminescence spectroscopy (HRPL) based on a scanning Fabry-P\'{e}rot
interferometer with a resolution better than 240~MHz ($1\,\mu$eV) (FWHM) has been applied~\cite{Ates.ea:2009,Ulrich.ea:2011,Ulhaq.ea:2012}.

%%%% Insert Fig. 1 here

By micro-photoluminescence measurements, we have pre-selected appropriate QDs with excitonic
ground state (s-shell) emission detuned by max.~$\pm 10$~GHz ($\pm 41\,\mu$eV) relative to the cesium $D_1$
absorption resonance $6^2S_{1/2} \rightarrow 6^2P_{1/2}$ at $\sim 335.11605$~THz/1.385928~eV %(894.59296~nm)
(excluding hyper-fine structure). For absorption measurements by using narrow-band RF of a single dot, this
restriction on the bare s-shell detuning from Cs-$D_1$ is of importance for the achievable resolution,
as dressed QD RF is affected by spectral broadening $\propto \Omega_0^2$ (with the bare Rabi frequency
$\Omega_0 \propto (P_0)^{1/2}$) with increasing resonant excitation power $P_0$ \cite{Ulrich.ea:2011,Roy.Hughes:2011}.
In addition, the QD dressed state emission line width depends on laser detuning $\Delta$ itself
and is explicitly ruled by the ratio of pure and radiative dephasing \cite{Ulhaq.ea:2013}.

\begin{figure}[!h]
 \includegraphics[width=0.5\textwidth]{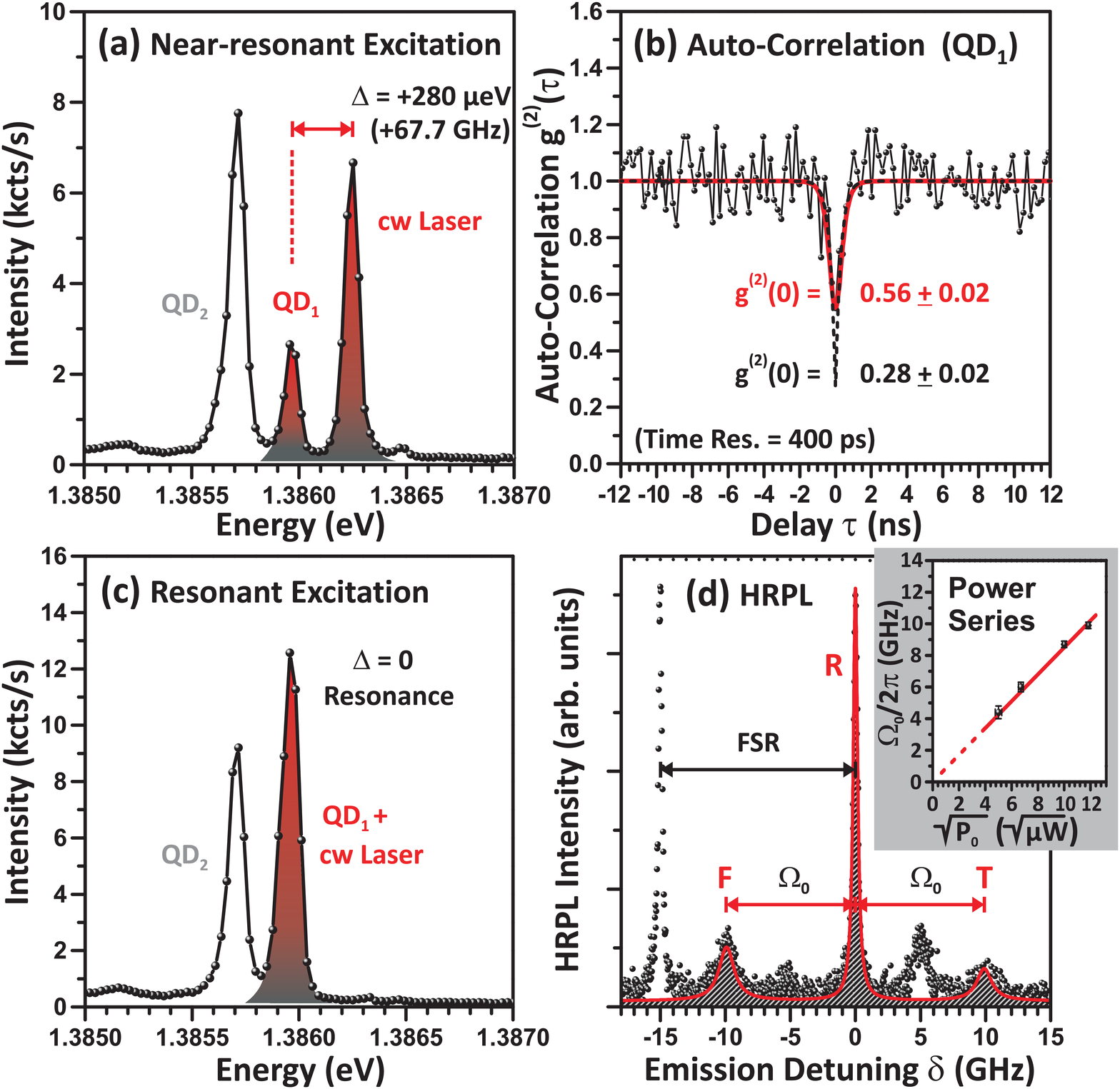}
 \caption{}
 \label{fig:1}
\end{figure}

Figure \ref{fig:1}(a) depicts the near-resonantly excited spectrum of a dot ($QD_1$)
which reveals excitonic emission at 335.12609~THz (1.38597~eV). While the laser is slightly blue-shifted
from the s-shell by $\Delta = +67.7$~GHz ($+280\,\mu$eV) and not in resonance with any
of its excited electronic states, selective pumping of $QD_1$ (accompanied by another nearby dot $QD_2$ with similar s-shell
resonance) is the result of incoherent excitation via carrier scattering with the bath of acoustic
phonons in the barrier~\cite{Weiler.ea:2012}. All measurements described in the following have been based on $QD_1$.

To investigate the quality of single-photon emission from $QD_1$, a continuous-wave $g^{(2)}(\tau)$
second-order auto-correlation has been performed on the spectrally filtered emission in (a). Figure~\ref{fig:1}(b)
shows the Poisson-normalized trace of the data, together with an analytic fit for the ideal correlation
$g^{(2)}(\tau) = 1 - \rho^2 \exp{(-|\tau|/\tau_p)}$ ($\tau_p^{-1}$: effective emitter re-pump rate), convolved with the
measured temporal response of our detection system, approximated by a Gaussian distribution of $\Delta \tau_{det} = 400$~ps
FWHM (red bold trace). Correcting for this temporal resolution $\Delta \tau_{det}$, we find a value of $g^{(2)}(0) = 0.28 \pm 0.02$,
representing a small contribution of uncorrelated background of only $\approx 15\,\%$ (mainly by overlap with $QD_2$).

The conditions of strictly resonant s-shell excitation of $QD_1$ ($P_0 = 150\,\mu$W; $\Delta = 0$) are shown in Fig.~\ref{fig:1}(c),
where the onset of resonance fluorescence reflects in an overall signal increase ($QD_1$ + laser) by factor $\sim 2$
with respect to near-resonant pumping in (a). In order to prove and investigate the onset of dressed state emission (unresolved in Fig.~\ref{fig:1}(c)),
a series of power-dependent HRPL measurements have been performed. Figure~\ref{fig:1}(d) depicts an example, revealing the
characteristic Mollow triplet~\cite{Mollow:1969} consisting of the central \textit{Rayleigh} peak ($R$)
and two symmetric side peaks denoted as the \textit{fluorescence} line ($F$) and \textit{three-photon} line ($T$), respectively.
At zero laser-QD detuning ($\Delta = 0$), the measured line splitting represents the \textit{bare Rabi frequency}
($\Omega_0/2\pi = 9.9 \pm 0.2$~GHz in this case), whereas $\Omega_0$ is replaced by an \textit{effective Rabi frequency}
$\Omega_\text{eff}(\Delta) \geq \Omega_0$ under additional laser detuning $|\Delta| > 0$ (see discussion of Fig.~\ref{fig:2} below).
The inset graph of Fig.~\ref{fig:1}(d) summarizes the evolution of the sideband splitting $\Omega_0/2\pi$
and confirms the theoretically expected linear dependence $\Omega_0 \propto (P_0)^{1/2}$
on the square root of excitation power~\cite{Mollow:1969}. From a Lorentzian fit to the Mollow triplet in Fig.~\ref{fig:1}(d),
we have derived very narrow line widths (FWHM) of $\Delta \nu_\text{F/T} = 1.3\,\pm\,0.2$~GHz for the $F/T$ lines.
Compared with the theoretical expectation of Fourier transform-limited inelastic emission bandwidth from
each triplet sideband of $\Delta \nu_\text{F/T}^\text{ideal} = 3/(4\pi T_1) \approx 0.3$~GHz (for a typical radiative
emitter lifetime of $T_1 = 800$~ps), the observed $F/T$ line widths are found to be slightly broadened as a result of pure
and predominantly excitation-induced dephasing~\cite{Ulhaq.ea:2013}. In contrast, the central Rayleigh $R$ line reveals
a very narrow (resolution-limited) line width of $0.4\,\pm\,0.2$~GHz. In principal, the \textit{pure} $R$ line in
the Mollow triplet dressed state regime should arise mainly from incoherently (but negligible contribution of coherently)
scattered RF signal. Though, reflected in an increased peak area ratio between $R$ and the $F/T$ side peaks of $\sim\,2.5$ (theory: 2)
in the experiment, the Rayleigh peak appears to be superimposed by residual laser stray light as the shape-dominating contribution.

%%%% Insert Fig. 2 here

The applied scheme of spectral resonance tuning between the pre-characterized $QD_1$ and the atomic $^{133}Cs$ vapor is depicted in Fig.~\ref{fig:2},
where the hyperfine structure of the relevant $D_1$ atomic transitions are shown in detail in (a), together with a sketch of the
theoretical dependence of RF dressed state emission on laser detuning $\Delta = \nu_\text{laser} - \nu_0^{QD}$ from the QD s-shell.

As is shown in Fig.~\ref{fig:2}(a), the $Cs-D_1$ absorption forms a quadruplet of dipole-allowed optical transitions, where the upper $6^2P_{1/2}$
state reveals a HFS of $\sim 1.1677$~GHz between $F = 3$ and 4. In addition, the $6^2S_{1/2}$ ground state HFS is $\sim 9.1926$~GHz~\cite{Steck:1998}.
At this point it is worth emphasizing first that both HFS values are of comparable magnitude to or even significantly larger than the observed emission
line widths $\Delta \nu$ of the Mollow triplet components of $QD_1$ (see Fig.~\ref{fig:1}(d)), as an important precondition for spectral line separation in absorption.
Moreover, the calculated transmission profile of $Cs$ vapor at $T = 293$~K ($20^\circ C$) and 7.5~cm optical absorption
length reveals Doppler-broadened line widths of $\Delta \nu_{FF'} \sim 0.45-0.50$~GHz (with transition index $F, F' = 3,4$) which are similar or only somewhat
smaller than the $R$, $F$, and $T$ probe light FWHM values discussed above.

Figure~\ref{fig:2}(b) sketches the principle of resonance tuning between RF from a single QD and the $D_1$ HFS of cesium. Whereas the $D_1$ quadruplet
is used as a fixed reference frame of absorption channels, we apply instantaneous spectral tuning of all three Mollow triplet components by precise detuning
of the laser frequency against the bare QD s-shell resonance. According to the theory of dressed state emission~\cite{Vamivakas.ea:2009,Ulhaq.ea:2013}, the frequencies of the triplet
components $R$ and $F/T$ have significantly different dependence on the explicit value of laser-QD detuning $\Delta$. Whereas the center Rayleigh line ($R$) reveals
a linear shift with the laser frequency, i.e. $\nu_R(\Delta) = \nu_0^{QD} + \Delta$, the RF side bands obey non-linear tuning as
$\nu_F(\Delta) = \nu_0^{QD} + \Delta - \Omega_\text{eff}(\Delta)/2\pi$ and $\nu_T(\Delta) = \nu_0^{QD} + \Delta + \Omega_\text{eff}(\Delta)/2\pi$
with the $\Delta$-dependent \textit{effective Rabi frequency} $\Omega_\text{eff} = (\Omega_0^2 + \Delta^2)^{1/2}$ and $\Omega_\text{eff}(\Delta) \geq \Omega_0$.
Worth to emphasize, the optical transitions via the $F$ or $T$ decay channels of the Mollow triplet represent sources of background-free single-photons
as was demonstrated in recent experiments \cite{Weiler.ea:2013}.

%%%% Insert Fig. 3 here

Using the full RF Mollow triplet emission of $QD_1$ as a source of narrow-band and tunable probe light, we have performed absorption series on atomic $Cs$.
A quartz cell with $Cs$ vapor (length: 7.5~cm; $T = 293$~K) was inserted into the detection path between the monochromator ($\approx\,35$~GHz bandpass for the RF) and an avalanche photo-diode (APD), connected to a digital rate meter for monitoring the transmission signal in dependence on laser detuning $\Delta$. For the scan, the laser power was stabilized
to $P_0 = 120 \pm 10\,\mu$W, equivalent to a constant bare Rabi frequency/splitting $\Omega_0 = 9.0 \pm 0.2$~GHz at $\Delta = 0$ as the starting conditions.
The cw Ti:Sapphire pump laser on $QD_1$ was manually frequency-scanned in steps of $0.2-0.3$~GHz (resolution-limited by the monitoring wave meter), and the according
averaged APD count rate was synchronously recorded. Figure~\ref{fig:3}(a) shows an excerpt of the full scan series taken between $\nu_\text{laser} = 335.1166$~THz
($\Delta = -9.6$~GHz) and $\nu_\text{laser} = 335.1289$~THz ($\Delta = +2.7$~GHz). As can be clearly seen from the transmission trace (black dots), we observe a series of distinct, narrow
transmission minima in the detuning range of $\Delta \approx -3$~GHz to $-6$~GHz, with $|\Delta T|/T = 25-40\,\%$ contrast well above the signal noise level ($\pm\,500$~cts/s,
equivalent to $\pm\,(3-4)\%$) and slight residual oscillations of the laser power apparent far-off resonance.

In order to analyze and interpret the absorption spectrum Fig.~\ref{fig:3}(a) with respect to distinct resonances between QD RF and the $D_1$ quadruplet,
the individual $\Delta$-dependent scalings of lines $F$, $R$, and $T$ have been calculated, as shown in Fig.~\ref{fig:3}(c). As stated earlier,
the bare Rabi frequency $\Omega_0/2\pi = 9.0 \pm 0.2$~GHz was fixed in the experiment, in accordance with the power series in Fig.~\ref{fig:1}(d) (inset).
The frequencies $\nu_{FF'}$ ($F,F' = 3,4$) denoting the $D_1$ transitions between $F = 3 \rightarrow 3$, $F = 3 \rightarrow 4$, $F = 4 \rightarrow 3$, and $F = 4 \leftrightarrow 4$
quantum states are represented as horizontal reference lines in the graph.

For a full theoretical analysis, we have calculated the expected transmission spectra of lines $F$, $R$, and $T$ through the atomic
vapor cell ($T = 293$~K), depicted in Fig.~\ref{fig:3}(b). In this evaluation, starting with each single transmission profile
(plotted as thin black (dashed/dotted) traces), the RF line widths and frequency scaling with $\Delta$ have been taken into account,
together with the Doppler broadening of the atomic transitions, resulting in Voigt-type profiles after convolution \cite{Siddons.ea:2008}.
The full transmission spectrum is shown as the bold-red trace in Fig.~\ref{fig:3}(b) and represents the superposition of all contributions, weighted by the
respective relative strengths of the Mollow triplet components derived from our HRPL spectra (see Fig.~\ref{fig:1}(d)). The only free fit parameter to the experimental data
is the average detector count rate of the transmitted light level (here: $16.0 \pm 0.2$~kcts/s) for detunings $\Delta$ far-off resonance. With this we directly find
very good agreement between theory and experiment, as can be seen from the superposition of the scaled theory spectrum (red-bold) and the experimental trace in (a).

As becomes obvious from a direct comparison between Figs.~\ref{fig:3}(a)-(c), we can identify the
absorption dips in (a) as resonances between the \textit{Rayleigh} ($R$) line and the $\nu_{34}$ (orange) and $\nu_{33}$ (red) transitions, as well as
resonances of the lower energetic \textit{fluorescence} line $F$ with $\nu_{44}$ (green) and $\nu_{43}$ (blue) of the atomic system, respectively. For the $T$ line though,
no such absorption resonances are expected within the experimental range of detuning $\Delta$. As a consequence of the selected fixed bare Rabi frequency
$\Omega_0/2\pi = 9.0$~GHz being very close to the overall splitting of 9.1926~GHz between the upper and lower doublet
of $\nu_{34}/\nu_{33}$ and $\nu_{44}/\nu_{43}$, all four resonance dips appear within a narrow window of $\Delta$. The $R-\nu_{33}$ resonance is well separated
by $\sim 0.9$~GHz and reveals distinct absorption contrast of $|\Delta T|/T \approx 25\,\%$. On the other hand, the remaining three resonances fall within a $\sim 1$~GHz
narrow band for the laser detuning. In accordance with theory (Fig.~\ref{fig:3}(b)), we can interpret the shape of this
absorption dip around $\Delta \approx -4.4$~GHz as a superposition, dominated by the $R-\nu_{34}$ signature. As a direct consequence of the relative weight of
spectral contributions by the strong $R$ line in comparison to the weaker $F$ line (ratio $\sim 2.5:1$, see above discussion and Fig.~\ref{fig:1}(d))
in combination with Doppler broadening, the two nearby transmission minima of $F-\nu_{43}$ and $F-\nu_{44}$ provide only small contributions
to the overall signal and remain unresolved. Worth to note, this close coincidence of absorption minima is not general, but only a consequence of our explicit
experimental starting conditions of initial QD s-shell detuning from the $D_1$ quadruplet in combination with the bare Rabi splitting chosen here.
Full separation of QD-atom resonances should be feasible by setting the dressing laser power such that $\Omega_0/2\pi \neq 9.1926$~GHz, i.e. distinctly different
from the HFS of the $Cs$ $6^2S_{1/2}$ ground state. \\

As an important remark already at this stage, we like to emphasize that single (In,Ga)As QD dressed-state resonance fluorescence (RF) is shown to provide highly
flexible tuning capability in conjunction with very narrow ($\leq 1$~GHz/$\sim 4\;\mu$eV range) emission line widths similar to the transitions of atomic $Cs$ vapor.
Therefore, this combined system represents a promising platform for sophisticated hybrid functionality between semiconductor QDs and atoms in future applications.
Among others one might anticipate, e.g. the filtering and/or stabilization of single photon emission~\cite{Akopian.ea.arXiv:2013} with respect to the frequency standard
of $Cs$-based atomic clocks, enhanced efficiency generation of slow light and storage/retrieval operations of single photons in quantum networks (repeaters) \cite{Reim.ea:2011,Choi.ea:2008},
or the non-linear controlled interaction of single photons with Rydberg atomic vapor \cite{Peyronel.ea:2012,Li.ea:2013}. In all the latter examples, the similarity between the QD emission profile
and the absorption/transition spectra of the interacting thermal atomic vapor will play a crucial role, enabling to severely enhance any non-linear optical response of the atoms.

In conclusion, we have demonstrated controllable hybrid resonance tuning and coupling between a semiconductor quantum emitter and alkali-atomic gas. In our
scheme, the precisely tunable dressed state resonance fluorescence (RF) from a single (In,Ga)As/GaAs quantum dot served as narrow band probe light to optically map
the $D_1$ hyperfine structure of cesium ($^{133}Cs$) vapor around $\sim 335.116$~THz ($\sim 894.592$~nm). Distinct resonances between individual QD Mollow triplet lines
and the atomic transitions were clearly identified from absorption spectra. Our findings and interpretations could be fully verified by theoretical calculations
on the expected transmission spectra under consideration of the given experimental conditions. These results show that narrow-band, optically dressed QD emission is
a highly promising and versatile alternative to competing hybrid tuning schemes by application of magnetic or electric fields.

The authors gratefully acknowledge financial support by the Deutsche Forschungs-Gesellschaft (DFG) through project DFG~MI~500/23-1. S. W. acknowledges funding by
the Carl-Zeiss-Stiftung.

\bibliography{text}% Produces the bibliography via BibTeX.

\newpage

\newpage

\begin{figure}[!h]
 \includegraphics[width=0.47\textwidth]{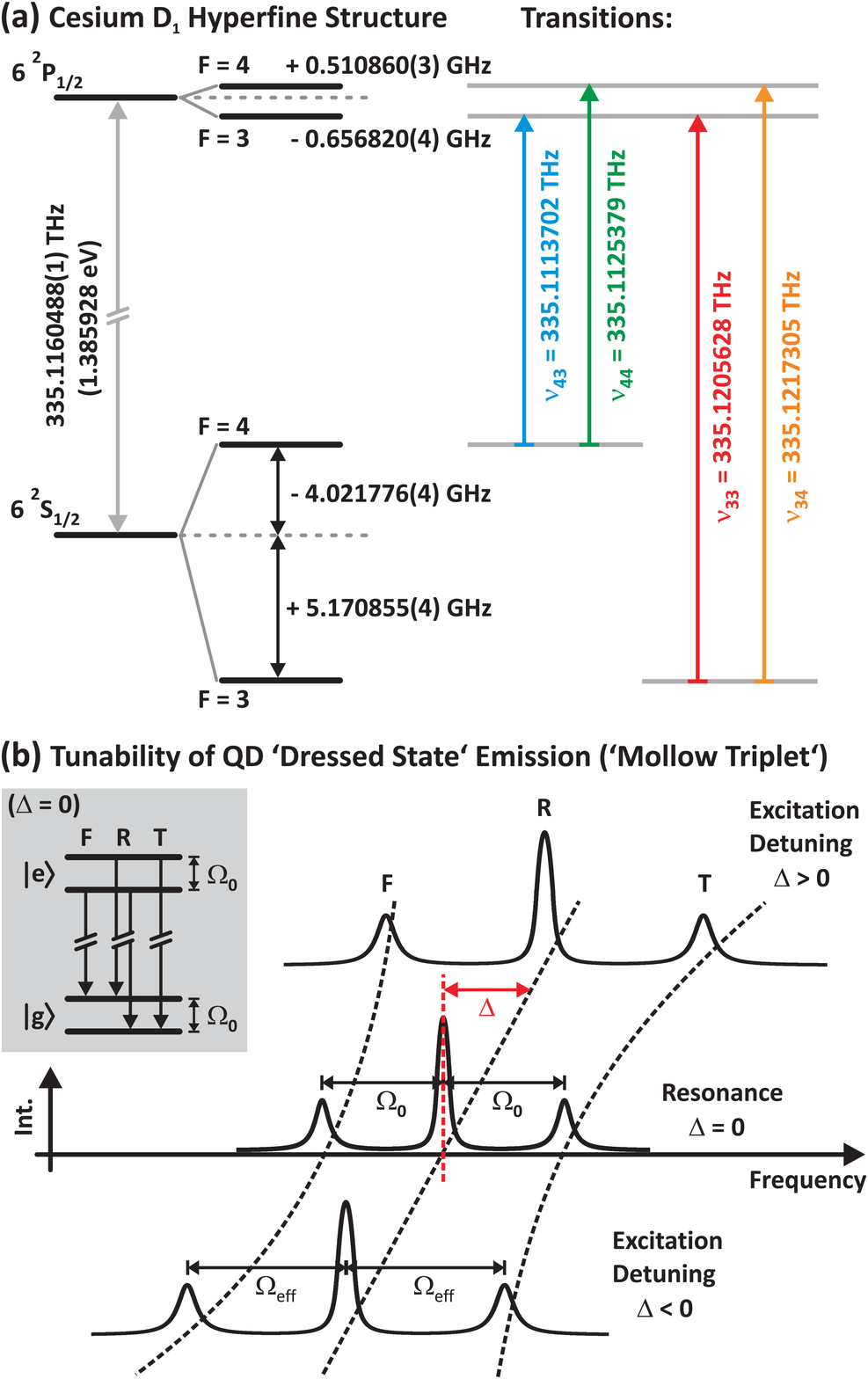}
 \caption{}
 \label{fig:2}
\end{figure}

\newpage

\begin{figure}[!h]
\includegraphics[width=0.49\textwidth]{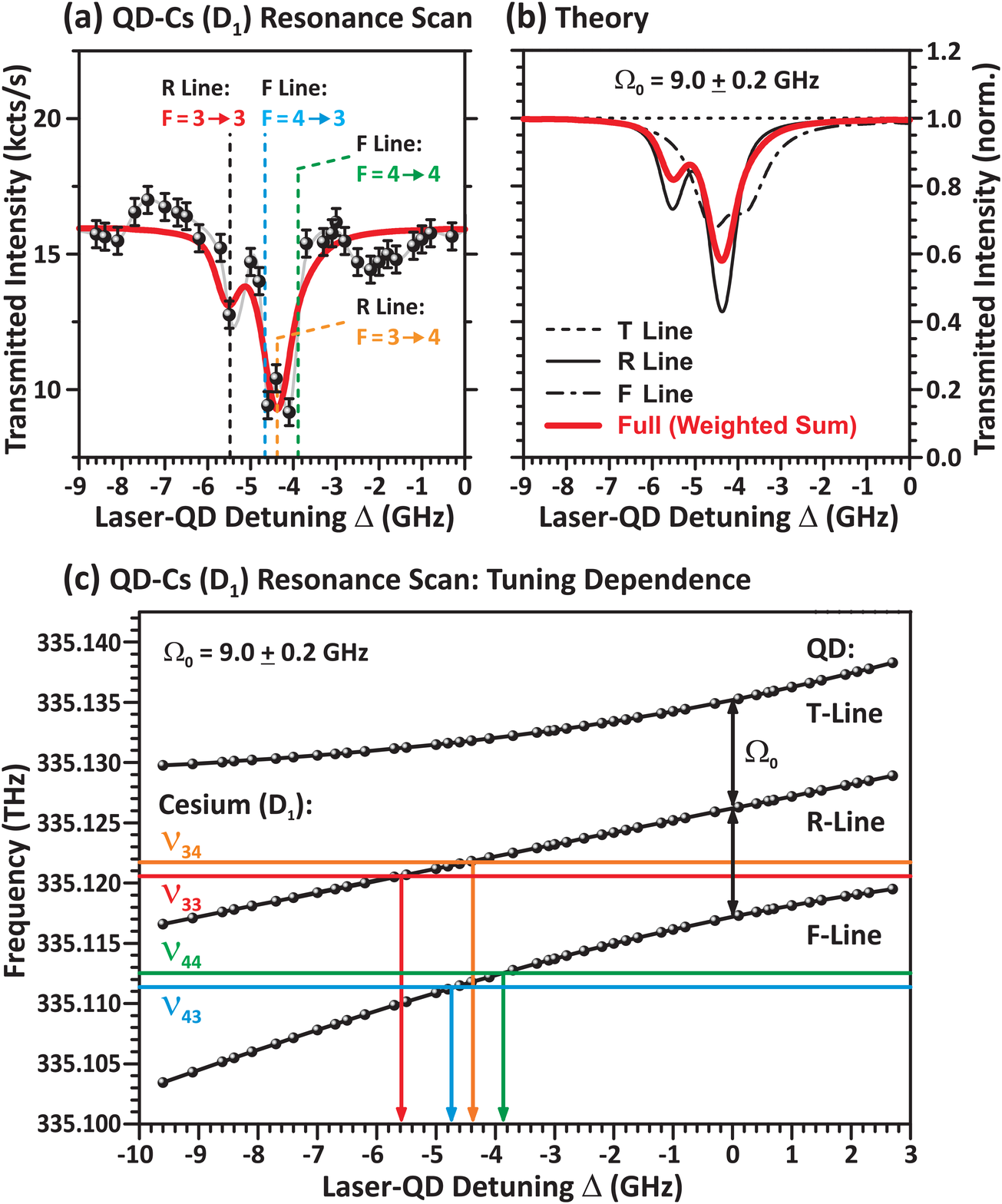}
\caption{}
\label{fig:3}
\end{figure}

\newpage

\textbf{FIGURE 1 - Caption}

\textbf{Fig.~1} (a) Low-temperature ($T = 4$~K) emission spectra of two
(In,Ga)As/GaAs quantum dots selectively excited under near-resonant optical
pumping (Laser-QD$_1$ detuning $\Delta = +280\,\mu\text{eV}\;/\;67.7$~GHz;
$P_0 = 150\,\mu$W).  (b) Second-order correlation of the spectrally
filtered QD$_1$ excitonic emission line in (a). The bold (dotted) line
corresponds to a fit to the raw data convoluted (deconvoluted) with the
temporal detector response ($\Delta \tau_\text{det} = 400$~ps).
The $\Delta \tau_\text{det}$-corrected value of $g^{(2)}(0) = 0.28\,\pm\,0.02$
corresponds to only $\approx 15\%$ uncorrelated background by overlap
with QD$_2$.  (c) Strictly resonant emission spectrum ($\Delta = 0$;
$P_0 = 150\,\mu$W) of QD$_1$, revealing the onset of RF from a signal
increase by factor $\sim 2$ in comparison with near-resonant excitation (a).
(d) Corresponding high-resolution (HRPL) measurement of the dressed state RF
of QD$_1$, revealing the characteristic Mollow triplet and a bare Rabi
splitting of $\Omega_0 = 9.9\,\pm\,0.2$~GHz in this case. Inset: Linear dependence
of the Rabi splitting with square root of power $\Omega_0 \propto (P_0)^{1/2}$
in accordance with theory~\cite{Mollow:1969}.

\newpage

\textbf{FIGURE 2 - Caption}

\textbf{Fig.~2} (a) Diagram of the cesium-$D_1$ transition between states
$6^2S_{1/2}$ and $6^2P_{1/2}$. A quadruplet of dipole-allowed transitions
is the result of electron-nucleus spin hyperfine interaction, denoted by
states with total atomic angular momenta $F = 3$ and $F = 4$~\cite{Steck:1998}.
The upper and lower states are split by $\sim 1.1677$~ GHz and $\sim 9.1926$~GHz,
respectively. (b) Detuning dependence of dressed state (Mollow triplet) emission
on laser detuning $\Delta = \nu_\text{laser} - \nu_0^{QD}$ between the excitation
laser and the bare resonance (QD: s-shell transition) of the involved two-level emitter system.
The position and Rabi splitting $\Omega_\text{eff}(\Delta)$ of lines $R$ and
$F/T$ can be tuned by the value of $\Delta$ to achieve resonance with the $Cs-D_1$
quadruplet, as discussed in the text. Inset: Level diagram of the dressed
two-level emitter state at $\Delta = 0$.~\cite{Mollow:1969}

\newpage

\textbf{FIGURE 3 - Caption}

\textbf{Fig.~3} (a) Resonance scan between the RF (Mollow triplet) of
$QD_1$, showing the photon signal after passing a quartz cell with $^{133}Cs$
vapor. In the experiment, the laser power was stabilized to $P_0 = 120\,\mu$W,
equivalent to a bare Rabi frequency/splitting $\Omega_0/2\pi = 9.0 \pm 0.2$~GHz.
The clear dips in the absorption spectrum (black dots) can be identified as resonances
between the $R$ and $F$ resonance fluorescence lines and the $Cs-D_1$ quadruplet.
(b) Calculated Doppler-broadened absorption spectra of QD RF after passing
a 7.5~cm long cell with atomic $Cs$ vapor at $T = 293$~K. Normalized
absorption spectra of the individual $F$, $R$, and $T$ bands are
depicted in black, together with the full signal (bold red line), representing
their weighted sum. (c) Analysis of the absorption resonances between $QD_1$ and
$Cs-D_1$. Black symbols represent the spectral positions of the center ($R$)
and side bands ($F, T$) of the RF (Mollow triplet) in dependence on laser
detuning $\Delta = \nu_\text{laser} - \nu_0^{QD}$ from the QD s-shell.
Resonances between the QD RF ($R$ and $F$ line) and the $Cs-D_1$
transitions $\nu_{FF'}$, indicated by horizontal lines, can be clearly
identified and assigned to corresponding dips in the transmission signal (Fig.~\ref{fig:3}(a)).

\end{document}